\documentclass[twocolumn]{aastex61}

\received{}
\revised{}
\accepted{\today}
\submitjournal{ApJ}

\shorttitle{[Te~{\sc iii}] and [Br~{\sc v}] Emission Lines in Planetary Nebulae}
\shortauthors{Madonna et al.}

\newcommand{\te}{$T_{\rm e}$}

\newcommand{\fkriii}{[Kr~{\sc iii}]}

\newcommand{\fkrvi}{[Kr~{\sc vi}]}

\newcommand{\frbiv}{[Rb~{\sc iv}]}

\newcommand{\ffeii}{[Fe~{\sc ii}]}

\newcommand{\fseiii}{[Se~{\sc iii}]}
\newcommand{\fseiv}{[Se~{\sc iv}]}
\newcommand{\fbriii}{[Br~{\sc iii}]}
\newcommand{\fbriv}{[Br~{\sc iv}]}
\newcommand{\fbrv}{[Br~{\sc v}]}

\newcommand{\ciii}{C~{\sc iii}}

\newcommand{\fcdiv}{[Cd~{\sc iv}]}
\newcommand{\fgevi}{[Ge~{\sc vi}]}
\newcommand{\fteiii}{[Te~{\sc iii}]}
\newcommand{\teii}{Te~{\sc ii}}
\newcommand{\tei}{Te~{\sc i}}

\begin{document}


\title{Neutron-Capture elements in planetary nebulae: first detections of near-Infrared {\fteiii} and {\fbrv} emission lines\footnote{This paper includes observations made with the Gran Telescopio Canarias (GTC), installed in the Spanish Observatorio del Roque de los Muchachos of the Instituto de Astrofísica de Canarias, on the island of La Palma. Programme ID GTC8-17AESCIVER}\footnote{This paper includes data taken at The McDonald Observatory of The University of Texas at Austin.}}

\correspondingauthor{Simone Madonna}
\email{smadonna@iac.es}

\author{Simone Madonna}\affiliation{Instituto de Astrof\'{\i}sica de Canarias, E-38200, La Laguna, Tenerife, Spain}\affiliation{Universidad de La Laguna. Depart. de Astrof\'{\i}sica, E-38206, La Laguna, Tenerife, Spain} \author{Manuel Bautista}\affiliation{Department of Physics, Western Michigan University, Kalamazoo, MI 49008, USA} \author{Harriet L. Dinerstein}\affiliation{Department of Astronomy, University of Texas, 2515 Speedway, C1400, Austin, TX 78712-1205, USA} \author{N. C. Sterling}\affiliation{Department of Physics, University of West Georgia, 1601 Maple Street, Carrollton, GA 30118, USA} \author{Jorge Garc\'{\i}a-Rojas}\affiliation{Instituto de Astrof\'{\i}sica de Canarias, E-38200, La Laguna, Tenerife, Spain}\affiliation{Universidad de La Laguna. Depart. de Astrof\'{\i}sica, E-38206, La Laguna, Tenerife, Spain} \author{Kyle F. Kaplan}\affiliation{Department of Astronomy, University of Texas, 2515 Speedway, C1400, Austin, TX 78712-1205, USA}\affiliation{Department of Astronomy \& Steward Observatory, 933 North Cherry Ave., University of Arizona, Tucson, AZ 85721-0065}\author{Maria del Mar Rubio-D\'{\i}ez}\affiliation{Centro de Astrobiolog\'{\i}a, , CSIC-INTA, Carretera de Torrej\'on a Ajalvir, km 4, E-28850 Torrej\'on de Ardoz, Spain} \author{Nieves Castro-Rodr\'{\i}guez}\affiliation{GRANTECAN, Cuesta de San Jos\'e s/n, E-38712 , Bre\~na Baja, La Palma, Spain}\affiliation{Instituto de Astrof\'{\i}sica de Canarias, E-38200, La Laguna, Tenerife, Spain}\affiliation{Universidad de La Laguna. Depart. de Astrof\'{\i}sica, E-38206, La Laguna, Tenerife, Spain} \author{Francisco Garz\'on}\affiliation{Instituto de Astrof\'{\i}sica de Canarias, E-38200, La Laguna, Tenerife, Spain}\affiliation{Universidad de La Laguna. Depart. de Astrof\'{\i}sica, E-38206, La Laguna, Tenerife, Spain}




\begin{abstract}

We have identified two new near-infrared emission lines in the spectra of planetary nebulae (PNe) arising from heavy elements produced by neutron capture reactions: {\fteiii} 2.1019 $\micron$ and {\fbrv} 1.6429 $\micron$. {\fteiii} was detected in both NGC\,7027 and IC\,418, while {\fbrv} was seen in NGC\,7027. The observations were obtained with the medium-resolution spectrograph EMIR on the 10.4m Gran Telescopio Canarias at La Palma, and with the high-resolution spectrograph IGRINS on the 2.7m Harlan J. Smith telescope at McDonald Observatory. New calculations of atomic data for these ions, specifically A-values and collision strengths, are presented and used to derive ionic abundances of Te$^{2+}$ and Br$^{4+}$. We also derive ionic abundances of other neutron-capture elements detected in the near-infrared spectra, and estimate total elemental abundances of Se, Br, Kr, Rb, and Te after correcting for unobserved ions. Comparison of our derived enrichments to theoretical predictions from AGB evolutionary models shows reasonable agreement for solar metallicity progenitor stars of $\sim$~2 -- 4 M$_{\odot}$. The spectrally-isolated {\fbrv} 1.6429 $\micron$ line has advantages for determining nebular Br abundances over optical {\fbriii} emission lines that can be blended with other features. Finally, measurements of Te are of special interest because this element lies beyond the first peak of the \emph{s}-process, and thus provides new leverage on the abundance pattern of trans-iron species produced by AGB stars.
\end{abstract}



\keywords{atomic data --- line: identification --- nuclear reactions, nucleosynthesis, abundances --- planetary nebulae: individual (IC 418)  --- planetary nebulae: individual (NGC 7027) --- stars: AGB and post-AGB}



\section{Introduction} \label{sec:intro}

Nucleosynthetic processes in asymptotic giant branch (AGB) stars produce a substantial fraction of the trans-iron (Z $>$ 30) elements present in the Solar System and the universe. 
During the late thermally-pulsing AGB evolutionary stage of low- and intermediate-mass ($\sim$1.5 -- 8 M$_{\odot}$) stars, 
iron peak nuclei in the intershell region between the He- and H-burning shells experience sequential neutron(\emph{n})-captures interlaced with $\beta$ decays, forming heavier elements via the so-called slow-neutron capture or ``\emph{s}-process''. The free neutrons are released as a result of alpha captures by $^{13}$C or $^{22}$Ne nuclei during phases of the thermal pulse cycle \citep{kappeleretal11}.
Convective dredge-up from the intershell region (third dredge up, hereafter TDU) may convey \emph{s}-processed material, along with C, to the stellar surface, which is  released into the interstellar medium by stellar winds and planetary nebula ejection \citep[and references therein]{karakaslattanzio14}.

Planetary nebulae (PNe) are valuable laboratories for studying the \emph{s}-process because they represent the final envelope compositions of their progenitor stars after the last thermal pulse and cessation of nucleosynthesis and TDU. The past few years have seen rapid development of near-infrared (NIR) spectroscopy as a tool for such investigations. The first NIR emission lines of \emph{n}-capture elements discovered, {\fkriii} 2.199 $\micron$ and {\fseiv} 2.287 $\micron$ \citep{dinerstein01}, have been widely observed in Milky Way and Magellanic Cloud PNe  (\citealp[][hereafter SPD15]{sterlingdinerstein08, sterlingetal15}; \citealp{mashburnetal16}). Additional NIR line detections include {\frbiv}, {\fcdiv}, {\fgevi}, {\fseiii} and {\fkrvi}  \citep{sterlingetal16, sterlingetal17}.

In this Letter we present the discovery of NIR lines from the \emph{n}-capture elements Br and Te, and atomic data needed to derive chemical abundances from the line intensities. Although optical lines of Te$^{2+}$ and Br$^{2+}$ have previously been reported in some PNe  (\citealp[hereafter SH07]{pequignotbaluteau94, sharpeeetal07}; \citealp[hereafter MGR17]{madonnaetal17}), they are faint and potentially blended with lines of other species. The discovery of relatively bright NIR lines of these  \emph{n}-capture elements provides an important new opportunity for determining the contributions of AGB stars to galactic enrichment. 

\begin{table}
\begin{minipage}{75mm}
\centering
\caption{Journal of observations.}
\label{tobs}
\begin{tabular}{c@{\hspace{2.8mm}}c@{\hspace{2.8mm}}c@{\hspace{2.8mm}}c@{\hspace{1.8mm}}c@{\hspace{1.8mm}}}
\hline
\hline
Instr.&$\Delta\lambda$~(\micron)  & $\lambda$/$\Delta\lambda$& \emph{t}$_{exp}$ (s) & Date  \\
\hline
\hline
NGC\,7027\\
\hline
EMIR& 1.17$-$1.33& 4800 & 720 &2016 Oct 18 \\
"& 1.52$-$1.77 & 4500 &720 & "    \\
"& 2.03$-$2.37 & 4000 & 720 & "  \\
IGRINS& 1.45$-$2.45 & 45000&1080 & 2014 Oct 24\\
\hline
IC\,418\\
\hline
EMIR&2.03$-$2.37 & 3500 &360 &  2017 Oct 5\\
IGRINS& 1.45$-$2.45 & 45000 &2400 & 2015 Nov 2\\
\noalign{\vskip3pt} \noalign{\hrule} \noalign{\vskip3pt}
\end{tabular}
\end{minipage}
\end{table}
\section{Observations and Data Reduction} \label{sec:obs}

\subsection{EMIR Observations} \label{sec:emir}

We observed the PNe IC\,418 and NGC\,7027 with the NIR spectrograph EMIR (Espectr\'ografo Multiobjeto Infra-Rojo) \citep{garzonetal06, garzonetal14} on the 10.4m Gran Telescopio Canarias at the Roque de los Muchachos Observatory, La Palma. 
The observations were carried out during commissioning and science verification runs of EMIR in longslit mode  (6.7$\arcsec \times$ 0.8$\arcsec$ for IC\,418 and 6.7$\arcsec \times$ 0.6$\arcsec$ for NGC\,7027), delivering spectral resolutions R $\sim$ 3500 in $K$ for IC\,418, and R $\sim$ 4800, 4500, and 4000 in $J$, $H$ and $K$ respectively for NGC\,7027. 
The slit was oriented N-S and centered on the central star of IC\,418. For NGC\,7027 the slit was centered on the eastern ridge at coordinates (RA, Dec, J2000) 21h 07m 02.16s, +42$\arcdeg$ 14$\arcmin$ 10.3 $\arcsec$, and was rotated to a position angle of 105$\arcdeg$ east of N-S to intersect the peak of the western ridge. 
We nodded 50$\arcsec$ along the slit for both objects to optimize observing efficiency. 
A0V standard stars were observed and used for spectrophotometric flux calibration for each PN.  We also used the standard star to correct for telluric absorption for IC\,418, but for NGC\,7027 we used a synthetic atmosphere spectrum calculated with ATRAN tools\footnote{\url{https://atran.sofia.usra.edu/cgi-bin/atran/atran.cgi}} due to a large airmass difference between the PN and its standard star. Wavelength calibration was performed using HgAr lamp spectra. The basic data reduction 
was carried out with {\sc iraf}, while we used MarTell (Rubio-D\'{\i}ez et al., in prep.) 
to perform the telluric absorption correction. Line fluxes were measured using the \emph{splot} routine in {\sc iraf} by integrating the flux above the local continuum. 
Table~\ref{tobs} presents a journal of the observations. 

\subsection{IGRINS Observations} \label{sec:igrins}

Both targets were also observed with the Immersion GRating Infrared Spectrometer (IGRINS) on the Cassegrain focus on the 2.7 m Harlan J. Smith Telescope at McDonald Observatory (Table~\ref{tobs}).  IGRINS is a NIR cross-dispersed echelle spectrometer that uses a silicon immersion grating to achieve high spectral resolving power, R $\sim$ 45000, with instantaneous complete coverage from 1.45 -- 2.45 $\micron$ \citep{parketal14}. The fixed long slit has dimensions of 1$\arcsec$ $\times$ 15$\arcsec$ on the sky and was rotated to a position angle 60$\arcdeg$ east of N-S for NGC\,7027, so that it intersected the peaks of both the eastern and western ridges. For IC\,418, the slit was oriented N-S and centered 3.5\arcsec\ west of the central star.
All observations were taken in a nod-off-slit pattern with sky frames at least 30$\arcsec$ from the source position. Standard A0V stars observed at similar airmasses as the PNe were employed for telluric correction and spectrophotometric flux calibration. The data reduction and analysis were performed as described in \cite{kaplanetal17}.

\begin{figure*}
 \centering
	\includegraphics[width=16cm]{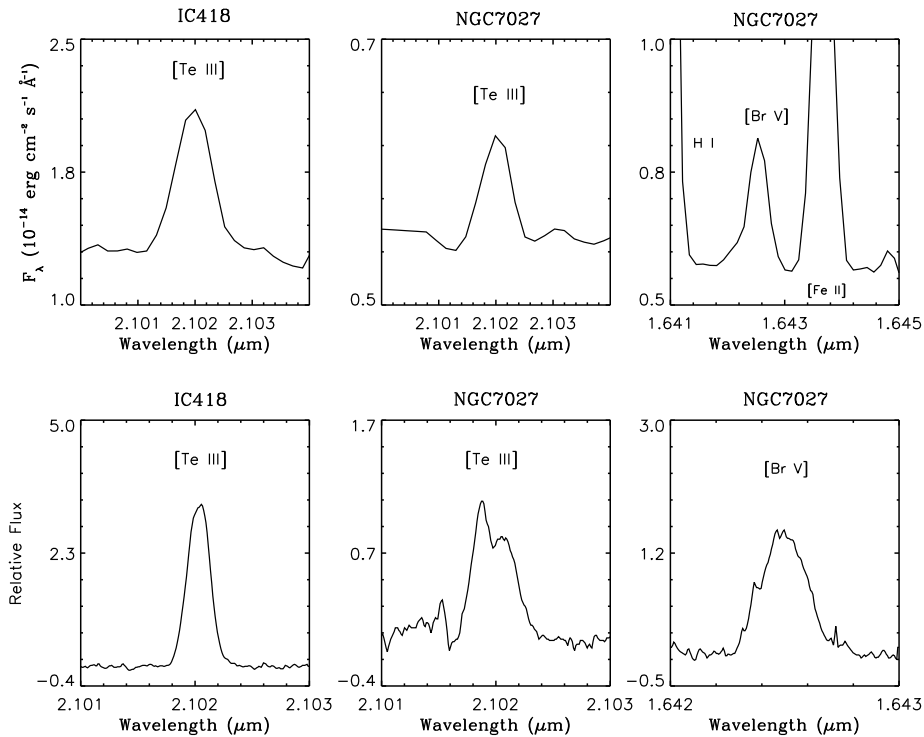}
    \caption{Plots of the newly detected lines from the EMIR (upper row) and IGRINS (lower row) data. While the EMIR line fluxes were measured from their 1D spectra, the IGRINS fluxes were measured in 2D position-velocity space with an aperture drawn around the line profile \citep[e.g. see Fig.\ 1 of][]{sterlingetal16}.}
    \label{lineprofile}
\end{figure*}

\section{Identification of {\fteiii} 2.1019 $\micron$ and {\fbrv} 1.6429 $\micron$} \label{sec:id}

Table~\ref{ion_abu} presents the intensities of NIR \emph{n}-capture element lines as ratios to nearby H~I lines, and
Fig.~\ref{lineprofile} shows excerpts of the spectral regions near the features we identify as {\fteiii} and {\fbrv}. While the lines are unresolved in the EMIR data, their profiles as seen with IGRINS reflect actual velocity structure and are consistent with the profiles of other ionic lines. Since the ionization potential (IP) of Te$^{+}$ is 18.55 eV and that of Te$^{2+}$ is 27.96 eV, a significant fraction of the Te in PNe with cool central stars, such as IC 418, should be in the form of Te$^{2+}$.
Despite its hotter central star, the bright, \emph{s}-process enriched PN NGC 7027 also shows {\fteiii}. Since stellar photons of at least 47.24 eV are required to produce Br$^{4+}$, it is not surprising that we do not see {\fbrv} in IC 418.

We identify the line near 2.1019 $\micron$ as the 5s$^{2}$5p$^{2}$ $^{3}$P$_{1}$ -- $^{3}$P$_{0}$ transition of Te$^{2+}$. Although the most recent reported vacuum laboratory value is 2.1021 $\micron$ \citep{tauheednaz11}, the IGRINS observations indicate that the actual wavelength is 2.1019 $\micron$, in good accord with the experimental data of \citet{joshietal92}.  The older energy level values of \citet{moore57} correspond to a wavelength of 2.1048~\micron.  This highlights the importance of ongoing atomic physics research, especially laboratory spectroscopy, for identifying \emph{n}-capture element transitions, as well the value of high-spectral resolution astronomical observations for measuring precise wavelengths.

We considered alternate identifications within $\pm$10\text{\AA} of the detected features
by querying the Atomic Line List v2.05b21\footnote{\url{http://www.pa.uky.edu/~peter/newpage/}}.  One of the most plausible alternate identifications is {\ffeii} 2.1017 $\micron$. However, other {\ffeii} lines of the same multiplet are not detected, although they are predicted to be brighter than this transition according to calculations with PyNeb v1.0.26 \citep{luridianaetal15}. 
Additionally, the feature at 2.1017 $\micron$ is too strong by several orders of magnitude to be consistent with the Fe$^{+}$ ionic abundance derived from {\ffeii} 1.2570, 12946, and 1.6440 $\micron$. An even closer wavelength coincidence with the observed feature is 4s $^{2}$P$_{1/2}$ -- $^{4}$P$_{1/2}$ [Ni~{\sc ii}] 2.1019 $\micron$, one of several transitions that can be enhanced over the strength expected for pure collisional excitation by UV continuum fluorescence  \citep{bautista96}. We eliminated this possibility since several other [Ni~{\sc ii}] transitions expected to be strong under such excitation conditions (1.7249, 1.7651, 2.0492, and 2.0811 $\micron$) are not detected.

Additional supporting evidence for the identification of the 2.1019 $\micron$ line as {\fteiii} is the consistency of Te$^{2+}$ ionic abundances derived from this line 
with values based on {\fteiii} $\lambda$7933.31. From the flux of this line in IC\,418 reported by 
\citet{sharpeeetal03} we find a Te$^{2+}$ abundance within $\sim$ 0.2 dex of that from the NIR line (see Table~\ref{ion_abu}).  
There is only an upper limit on the intensity of {\fteiii} $\lambda$7933.31 in NGC\,7027 (SH07), but this is consistent with our results since the corresponding 
limit on the Te$^{2+}$ abundance is larger than the value we derive from {\fteiii} 2.1019 $\micron$ (Table~\ref{ion_abu}).

We identify the line at 1.6429 $\micron$ seen in NGC\,7027 as the 4s$^{2}$4p $^{2}$P$^{o}$$_{3/2}$ -- $^{2}$P$^{o}$$_{1/2}$ transition of Br$^{4+}$, for which the energy levels provided by \cite{riyazetal14} yield a vacuum wavelength of 1.6429 $\micron$.
We searched the Atomic Line List and found no plausible alternate atomic lines. H$_2$~6-4 $Q(5)$, which can be strong in sources with highly fluorescent H$_2$ spectra (Kaplan, et~al., in preparation), is a potential contaminant, but the IGRINS data demonstrate that it is not present at a detectable intensity. 

The good agreement between EMIR and IGRINS results and the high resolution of IGRINS data, which eliminates the possibility of blends, give us confidence that 
 {\fteiii} and {\fbrv} respectively are the correct 
identifications for the 2.1019 $\micron$ and 1.6429 $\micron$ lines.

\section{A-values and Collision strengths}

The atomic structures, $A$-values, and radial wavefunctions for Br~{\sc v} and Te~{\sc iii} were computed with the 
{\sc autostructure} code \citep{badnell86, badnell11}. We diagonalized the Breit--Pauli Hamiltonian within a statistical Thomas--Fermi--Dirac--Amaldi model potential $V(\lambda_{nl})$ \citep{bad97}. 
The potential for each orbital was characterized by scaling the radial parameter by a quantity $\lambda_{nl}$ that is optimized variationally by minimizing
a weighted sum of the $LS$ term energies. The $LS$ terms are represented by configuration-interaction (CI) wavefunctions.

The CI expansion for the Br~{\sc v} system that we used includes the $3d^{10}4s^24p$, $3d^{10}4s4p^2$, $3d^{10}4s^25s$, $3d^{10}4s^25p$, $3d^{10}4s^25d$, $3d^94s^24p^2$, $3d^94s^24p4d$, and $3d^94s^24p5s$ configurations, while that for Te~{\sc iii} includes $4d^{10}5s^25p^2$,  $4d^{10}5s5p^3$, $4d^{10}5s^25p5d$, and $4d^{9}5s^25p^3$.

The quality of our atomic structure representations was evaluated by comparing predicted term energies with the experimental values of \cite{riyazetal14} and \cite{joshietal92} for Br~{\sc v} and Te~{\sc iii} respectively. Our calculated energies are found to differ from experimental values by less than 5\%. Additional small semi-empirical corrections to the orbitals were calculated through coupling of energy terms that minimized the relative energy differences. Finally, the level energies were shifted to the experimental values in order to compute  monopole, dipole, and quadrupole transition probabilities. 

The scattering calculations were done with the BP $R$-matrix program \citep{rmatrx}, using the orbitals from our {\sc autostructure} calculation,  retaining CI from 39 LS terms and 88 fine structure levels for Br~{\sc v} and  39 LS terms and 75 levels for Te~{\sc iii}. The calculations explicitly include partial waves from states with $L\le 9$ and multiplicities 1, 3, and 5 for Br~{\sc v} and 2 and 4 for Te~{\sc iii}. Contributions from the higher partial waves were estimated using a top-up procedure. 

Maxwellian-averaged collision strengths were computed for both ions for selected temperatures between 5000~K and 30,000~K. New atomic data files with the resulting values for Br~{\sc v} and Te~{\sc iii} were added to PyNeb. The $A$-values and Maxwellian-averaged collision strengths are presented in Table~\ref{atomic_data_table}.

\begin{table*}
 \scriptsize
 \caption{A-values and Collision Strenghts.}
  \label{atomic_data_table}
 \centering
 \begin{tabular}{cccccccc}
 \hline
 \hline
 & & & & $\Upsilon$(\emph{T}) &  & &\\
  \cline{3-8}
  Trans. & A$_{ij}$[s$^{-1}$] & 5000 K & 8000 K & 10000 K & 15000 K & 20000 K & 30000 K  \\
  \hline
  & & & & Te$^{2+}$  & & & \\
  \hline
$^{3}$P$_{1}$-$^{3}$P$_{0}$  &1.194&5.812 	& 	4.965 	& 	4.553 	& 	3.848 	& 	3.4 	& 	2.848 	\\
$^{3}$P$_{2}$-$^{3}$P$_{0}$&1.209E-02&2.565 	& 	2.597 	& 	2.62 	& 	2.652 	& 	2.642 	& 	2.561 	\\
$^{1}$D$_{2}$-$^{3}$P$_{0}$&1.062E-04&1.08 	& 	1.1 	& 	1.103 	& 	1.082 	& 	1.045 	& 	0.9653 	\\
$^{1}$S$_{0}$-$^{3}$P$_{0}$&...&0.1534 	& 	0.159 	& 	0.1626 	& 	0.1694 	& 	0.1721 	& 	0.1696 	\\
$^{3}$P$_{2}$-$^{3}$P$_{1}$&0.52&8.74 	& 	8.487 	& 	8.403 	& 	8.255 	& 	8.085 	& 	7.684 	\\
$^{1}$D$_{2}$-$^{3}$P$_{1}$&5.141&4.317 	& 	4.43 	& 	4.463 	& 	4.421 	& 	4.297 	& 	4.004 	\\
$^{1}$S$_{0}$-$^{3}$P$_{1}$&6.797E+01&0.6048 	& 	0.632 	& 	0.6414 	& 	0.6502 	& 	0.6452 	& 	0.6158 	\\
$^{1}$D$_{2}$-$^{3}$P$_{2}$&5.266&8.606 	& 	8.86 	& 	8.924 	& 	8.896 	& 	8.749 	& 	8.368 	\\
$^{1}$S$_{0}$-$^{3}$P$_{2}$&7.098&1.053 	& 	1.177 	& 	1.222 	& 	1.271 	& 	1.273 	& 	1.228 	\\
$^{1}$S$_{0}$-$^{1}$D$_{2}$&7.902&4.021 	& 	3.949 	& 	3.915 	& 	3.855 	& 	3.818 	& 	3.763 	\\
      \hline
      & & & & Br$^{4+}$  & & & \\
      \hline
$^{2}$P$^{0}$$_{3/2}$-$^{2}$P$^{0}$$_{1/2}$& 0.991	& 	5.39 	& 	5.014 	& 	4.908 	& 	4.892 	& 	5.055 	& 	5.507 	\\
$^{2}$D$_{3/2}$-$^{2}$P$^{0}$$_{1/2}$& 1.507E+06	& 	0.932 	& 	0.771 	& 	0.7088 	& 	0.6186 	& 	0.568 	& 	0.5084 	\\
$^{2}$D$_{5/2}$-$^{2}$P$^{0}$$_{1/2}$& 4.920E+04	& 	0.8377 	& 	0.7848 	& 	0.7679 	& 	0.7479 	& 	0.7355 	& 	0.7109 	\\
$^{2}$P$_{1/2}$-$^{2}$P$^{0}$$_{1/2}$& ...	& 	0.8529 	& 	0.8508 	& 	0.8575 	& 	0.8684 	& 	0.8653 	& 	0.8371 	\\
$^{2}$P$_{3/2}$-$^{2}$P$^{0}$$_{1/2}$& 7.741E+08	& 	3.292 	& 	3.208 	& 	3.179 	& 	3.138 	& 	3.112 	& 	3.069 	\\
$^{2}$D$^{0}$$_{3/2}$-$^{2}$P$^{0}$$_{1/2}$& ...	& 	1.505 	& 	1.585 	& 	1.633 	& 	1.708 	& 	1.733 	& 	1.707 	\\
$^{2}$D$^{0}$$_{5/2}$-$^{2}$P$^{0}$$_{1/2}$& 2.694E+09	& 	1.488 	& 	1.446 	& 	1.423 	& 	1.385 	& 	1.368 	& 	1.359 	\\
$^{2}$D$_{3/2}$-$^{2}$P$^{0}$$_{3/2}$& 4.842E+05	& 	1.522 	& 	1.244 	& 	1.14 	& 	0.9867 	& 	0.8946 	& 	0.7781 	\\
$^{2}$D$_{5/2}$-$^{2}$P$^{0}$$_{3/2}$& 6.594E+05	& 	1.601 	& 	1.549 	& 	1.536 	& 	1.502 	& 	1.458 	& 	1.368 	\\
$^{2}$P$_{1/2}$-$^{2}$P$^{0}$$_{3/2}$& 2.146E+06	& 	2.128 	& 	2.291 	& 	2.336 	& 	2.341 	& 	2.287 	& 	2.152 	\\
$^{2}$P$_{3/2}$-$^{2}$P$^{0}$$_{3/2}$& 6.632E+07	& 	2.793 	& 	2.618 	& 	2.571 	& 	2.514 	& 	2.476 	& 	2.397 	\\
$^{2}$D$^{0}$$_{3/2}$-$^{2}$P$^{0}$$_{3/2}$& 7.471E+08	& 	7.551 	& 	7.202 	& 	7.062 	& 	6.833 	& 	6.68 	& 	6.461 	\\
$^{2}$D$^{0}$$_{5/2}$-$^{2}$P$^{0}$$_{3/2}$& 1.313E+09	& 	1.586 	& 	1.525 	& 	1.482 	& 	1.402 	& 	1.357 	& 	1.313 	\\
$^{2}$D$_{5/2}$-$^{2}$D$_{3/2}$& 0.129	& 	4.457 	& 	3.822 	& 	3.521 	& 	3.027 	& 	2.735 	& 	2.424 	\\
$^{2}$P$_{1/2}$-$^{2}$D$_{3/2}$& 2.285E-04	& 	2.666 	& 	2.34 	& 	2.215 	& 	2.026 	& 	1.912 	& 	1.785 	\\
$^{2}$P$_{3/2}$-$^{2}$D$_{3/2}$& 0.666	& 	0.6057 	& 	0.5456 	& 	0.5272 	& 	0.5109 	& 	0.5135 	& 	0.5339 	\\
$^{2}$D$^{0}$$_{3/2}$-$^{2}$D$_{3/2}$& 2.818E-03	& 	0.3996 	& 	0.3728 	& 	0.3685 	& 	0.3783 	& 	0.4003 	& 	0.4478 	\\
$^{2}$D$^{0}$$_{5/2}$-$^{2}$D$_{3/2}$& 9.839	& 	0.1252 	& 	0.1689 	& 	0.1908 	& 	0.2288 	& 	0.2531 	& 	0.282 	\\
$^{2}$P$_{1/2}$-$^{2}$D$_{5/2}$& 0.277	& 	7.721 	& 	6.645 	& 	6.211 	& 	5.539 	& 	5.147 	& 	4.744 	\\
$^{2}$P$_{3/2}$-$^{2}$D$_{5/2}$& 2.092	& 	0.7815 	& 	0.7616 	& 	0.763 	& 	0.7834 	& 	0.8142 	& 	0.8767 	\\
$^{2}$D$^{0}$$_{3/2}$-$^{2}$D$_{5/2}$& 1.531	& 	1.294 	& 	1.16 	& 	1.119 	& 	1.09 	& 	1.106 	& 	1.174 	\\
$^{2}$D$^{0}$$_{5/2}$-$^{2}$D$_{5/2}$& 38.414	& 	0.2605 	& 	0.3401 	& 	0.38 	& 	0.448 	& 	0.4907 	& 	0.5414 	\\
$^{2}$P$_{3/2}$-$^{2}$P$_{1/2}$& 0.780	& 	1.142 	& 	1.008 	& 	0.9629 	& 	0.9157 	& 	0.9126 	& 	0.9448 	\\
$^{2}$D$^{0}$$_{3/2}$-$^{2}$P$_{1/2}$& 5.646	& 	2.394 	& 	2.123 	& 	2.032 	& 	1.929 	& 	1.903 	& 	1.924 	\\
$^{2}$D$^{0}$$_{5/2}$-$^{2}$P$_{1/2}$& 1.190	& 	0.2343 	& 	0.3052 	& 	0.355 	& 	0.4641 	& 	0.547 	& 	0.6547 	\\
$^{2}$D$^{0}$$_{3/2}$-$^{2}$P$_{3/2}$& 0.0002	& 	3.196 	& 	3.405 	& 	3.512 	& 	3.693 	& 	3.796 	& 	3.879 	\\
$^{2}$D$^{0}$$_{5/2}$-$^{2}$P$_{3/2}$& 21.699	& 	1.437 	& 	1.402 	& 	1.383 	& 	1.351 	& 	1.334 	& 	1.323 	\\
$^{2}$D$^{0}$$_{5/2}$-$^{2}$D$^{0}$$_{3/2}$& 30.410	& 	2.151 	& 	2.047 	& 	1.995 	& 	1.919 	& 	1.885 	& 	1.864 	\\
\hline
 \end{tabular}
\end{table*}

\section{Elemental Abundances and Enrichments} \label{sec:disc}

We recalculated the physical conditions, and ionic and elemental abundances of O, S, and Ar that are used in our abundance analysis, using PyNeb and line fluxes from multi-wavelength studies of IC\,418 \citep{bernardsalasetal01, pottaschetal04} and NGC\,7027 \citep{zhangetal05}. 
Multi-wavelength observations provide access to multiple ions per element, which minimizes systematic uncertainties due to 
the contributions of unobserved ions.
We take the total elemental abundances of O in IC\,418, and of Ar and S in both PNe, to be equal to the sum of all observed ionic abundances. For the high-excitation PN NGC\,7027, we correct for the presence of unobserved O$^{4+}$ using the formula from \citet{delgadoingladaetal14}.
The atomic data used and the treatment of uncertainties are the same as \citet{garciarojasetal15} and MGR17. Column~5 of Table~\ref{ion_abu} gives the \emph{n}-capture element abundances derived using our recalculated physical conditions and light element abundances, while Column~6 lists the values found using physical conditions and light ion abundances from the literature. The two sets of \emph{n}-capture abundances agree to within 0.1~dex in all cases.

\begin{table*}
	\centering
	\scriptsize
	\caption{\emph{n}-capture ionic and elemental abundances.}
	\label{ion_abu}
	\begin{tabular}{lccccc} 
	       \hline
	       \hline
		Line ($\mu$m) Ratio & Flux Ratio & log(X$^{i+}$/H$^{+}$)+12 & ICF$^{a}$ & log(X/H)+12$^{a}$      &  log(X/H)+12$^{b}$ \\	
		 &  &  &  & This work     &   \\       
                 \hline			
	        \multicolumn{6}{c}{NGC\,7027  ({\te} = 12450$\pm$600 K,   n$_{e}$ = 50500$^{+25700}_{-14600}$ cm$^{-3}$)}  \\
		\hline
			      \multicolumn{6}{c}{EMIR} \\
{\fseiv} 2.2864/Br$\gamma$	&	 0.0919$\pm$0.0055	&	3.20$\pm$0.04 & 2.52$\pm$1.12 & 3.60$\pm$0.16 &  3.56$\pm$0.14 \\
{\fbrv}  1.6429/Br11	&	0.0295$\pm$0.0021 	&	2.23$\pm$0.05 & 2.80$\pm$0.56 & 2.67$\pm$0.06 &  2.72$\pm$0.06\\
{\fkriii} 2.1986/Br$\gamma$	&	0.0329$\pm$0.0023		&	3.40$\pm$0.04 & 1.32$\pm$0.25	&4.13$\pm$0.08 &  4.11$\pm$0.08\\
{\fkrvi} 1.2330/P$\beta$	&	0.0078$\pm$0.0005	&	2.55$\pm$0.04 & --  &  --&--\\
{\frbiv} 1.5973/Br11	&	0.0335$\pm$0.0023		&	2.86$\pm$0.05&1.58$\pm$0.33 &3.06$\pm$0.09 &  3.03$\pm$0.09 \\
{\fteiii} 2.1019/Br$\gamma$	&	0.0030$\pm$0.0008		&	1.50$\pm$0.11 & 17.28$\pm$2.94& 2.74$\pm$0.12&  2.64$\pm$0.12\\
\hline
			           \multicolumn{6}{c}{IGRINS} \\

{\fseiv} 2.2864/Br$\gamma$	&	 0.0835$\pm$0.0083	&	3.16$\pm$0.05 &2.52$\pm$1.12  &3.56$\pm$0.17  &  3.52$\pm$0.14\\
{\fbrv}  1.6429/Br11	&	0.0245$\pm$0.0049 	&	2.15$\pm$0.08 &2.80$\pm$0.56& 2.59$\pm$0.09 &  2.64$\pm$0.09\\
{\fkriii} 2.1986/Br$\gamma$	&	0.0307$\pm$0.0031		&	3.38$\pm$0.05& 1.32$\pm$0.25	&4.12$\pm$0.09 &  4.10$\pm$0.08\\
{\frbiv} 1.5973/Br11	&	0.0290$\pm$0.0029		&	2.80$\pm$0.06 &1.58$\pm$0.33  & 3.00$\pm$0.10 &  2.97$\pm$0.10 \\
{\fteiii} 2.1019/Br$\gamma$	&	0.0031$\pm$0.0003		&	1.51$\pm$0.05 &17.28$\pm$2.94 & 2.75$\pm$0.07 &  2.65$\pm$0.07\\
\hline
			           \multicolumn{6}{c}{Literature$^{c}$} \\
{\fteiii} 0.7933/H$\beta$	&	$<$0.002		&	$<$2.15 &--  & -- & -- \\
                \hline

              \multicolumn{6}{c}{IC418  ({\te} = 8670$\pm$250 K ,  n$_{e}$ = 13650$^{+8200}_{-3400}$ cm$^{-3}$)}     \\
                \hline
			       \multicolumn{6}{c}{EMIR} \\    
{\fseiv} 2.2864/Br$\gamma$	&	0.0019$\pm$0.0008		 & 1.72$\pm$0.15 & 5.81$\pm$2.14&2.48$\pm$0.19 &  2.46$\pm$0.18 \\
{\fkriii} 2.1986/Br$\gamma$	&	0.0258$\pm$0.0015		 & 3.51$\pm$0.03 & 1.82$\pm$0.32&3.77$\pm$0.06 &  3.77$\pm$0.06\\
{\fteiii} 2.1019/Br$\gamma$	&	0.0122$\pm$0.0011		 &2.29$\pm$0.04 & 2.58$\pm$0.31& 2.70$\pm$0.07 &  2.67$\pm$0.07\\
 \hline
                
			         \multicolumn{6}{c}{IGRINS} \\
{\fseiv} 2.2864/Br$\gamma$	&	0.0014$\pm$0.0001		 &  1.57$\pm$0.04 & 5.81$\pm$2.14 &2.33$\pm$0.14 & 2.31$\pm$0.14 \\
{\fkriii} 2.1986/Br$\gamma$	&	0.0262$\pm$0.0026		 &  3.51$\pm$0.04 & 1.82$\pm$0.32&3.77$\pm$0.07 & 3.77$\pm$0.07\\
{\fteiii} 2.1019/Br$\gamma$	&	0.0126$\pm$0.0013 		 & 2.31$\pm$0.04 & 2.58$\pm$0.31&2.72$\pm$0.07 & 2.69$\pm$0.07\\
\hline
			           \multicolumn{6}{c}{Literature$^{d}$} \\
{\fteiii} 0.7933/H$\beta$	&		0.0017	& 2.53 &-- &   -- &--\\

\hline
\multicolumn{6}{l}{${\rm ^a}$ Results from our abundance re-calculations. For details of the adopted ICFs see \S~\ref{sec:brte}.}\\ 
\multicolumn{6}{l}{${\rm ^b}$ Calculated using values for physical conditions and for ionic and total abundances of light elements}\\ 
\multicolumn{6}{l}{from the literature (see \S~\ref{sec:brte}). ${\rm ^c}$ \cite{sharpeeetal07}. ${\rm ^d}$ \cite{sharpeeetal03}.}
	\end{tabular}
\end{table*}

\subsection{Elemental Abundances of Te and Br} \label{sec:brte}

Determination of elemental from ionic abundances requires correction for unobserved ionization stages. A widely used approach for such corrections is to construct formulae that define ionization correction factors (ICFs), where the ICF is the inverse of the fractional abundance of the element in the observed ion(s).
ICF prescriptions are most reliable when based on photoionization models that account for the radiative and collisional processes that affect the ionization balance of each element \citep[][SPD15]{delgadoingladaetal14}, but such models require atomic data for photoionization and recombination processes.  When these data are unknown, as is the case for Br and Te ions, approximate ICFs can be estimated using coincidences in IPs with ions of more abundant elements.  However, these approximate ICFs are much more uncertain than those derived from photoionization models.

To convert the measured ionic into elemental abundances for NGC\,7027 and IC\,418, we assumed that Te/Ar = Te$^{2+}$/Ar$^{+}$ and Br/Ar = Br$^{4+}$/Ar$^{3+}$, based on similarities between the IP ranges of those ions.  Adopting these procedures, we find  Te abundances in both PNe that are higher than solar by a factor of $\sim$3.5, while Br is modestly if at all enhanced in NGC\,7027.

\subsection{Abundances of other \emph{n}-capture Elements} \label{sec:literat}

Elemental abundances of other \emph{n}-capture elements detected in the EMIR and IGRINS spectra of NGC\,7027 and IC\,418  are also given in Table~\ref{ion_abu}. 
For NGC\,7027, we calculate the Kr total abundance using the ICF value of  \cite{sterlingetal17} which incorporates lines of four ions. For IC 418, we used Eq. 1 of SPD15, which uses the S$^{2+}$/S ratio and is based on Kr$^{2+}$ only, but the higher ions should not be prevalent in this low-ionization PN. The Rb abundance in NGC\,7027 was calculated assuming Rb/O = Rb$^{3+}$/O$^{2+}$ as in \cite{sterlingetal16}. 
For both PNe we computed the Se abundance using the ICF given by Eq. 8 of SPD15.

In Column 2 of Table~\ref{ncap_enrich} we list the averaged values from the EMIR and IGRINS data for the total abundances of the \emph{n}-capture elements. 
Column 3 reports the enhancements of these abundances relative to the solar values of \cite{asplundetal09}. 
We adopt the meteoritic abundances for all but Kr, which is based on solar wind measurements and interpolation from other elements \citep[see \S3.9.5 of][]{asplundetal09}.

Column 4 summarizes previous determinations of the elemental enrichments based on NIR emission lines, which agree within uncertainties with the values reported here.  While \citet{sterlingetal16} compared their Rb abundance with the photospheric solar Rb value of 2.52, after adjusting to the meteoritic value of 2.36 their abundance corresponds to an enhancement of [Rb/H] = 0.70, in excellent agreement with the current result.  The Se abundance we find for IC\,418, nearly a factor of 10 below solar, is not consistent with the paradigm of self-enrichment by the \emph{s}-process. 
However, in this low-excitation PN, most of the Se atoms will be in the form of Se$^{2+}$, which we have not observed, making the correction to the Se$^{3+}$ abundance large and uncertain (Table~\ref{ion_abu}).  In addition, \citet{sterlingetal17} suggested that the currently employed Se/Se$^{3+}$ ICF may
break down in  low- to moderate-excitation PNe, highlighting the importance of measuring the J-band {\fseiii}~1.0994~\micron\ line first reported by these authors.

\begin{table}
        \centering
	\scriptsize
	\caption{\emph{n}-capture element enrichments.}
	\label{ncap_enrich}
	\begin{tabular}{cccc} 
	       \hline
	       \hline
		Element & Elem. Abund.$^{a}$ &   [X/H]$^{b}$  & [X/H]$^{c}$ \\
		\hline
		 \multicolumn{4}{c}{NGC\,7027}    \\	       
		Se         & 3.58$\pm$0.17             & 0.24$\pm$0.17   & 0.34$\pm$0.10  \\
		Br         & 2.64$\pm$0.08             & 0.10$\pm$0.10   &   \\
		Kr         & 4.13$\pm$0.09             & 0.88$\pm$0.11   & 0.85$\pm$0.04 \\
		Rb         & 3.03$\pm$0.10             & 0.67$\pm$0.10   & 0.70$\pm$0.20  \\
		Te         & 2.75$\pm$0.10           & 0.57$\pm$0.10     &   \\
                 \hline
                  \multicolumn{4}{c}{IC\,418} \\
                 Se         & 2.40$\pm$0.17            & -0.94$\pm$0.17  & $<-0.32$   \\
		Kr         & 3.77$\pm$0.07             & 0.52$\pm$0.09   & 0.50$\pm$0.20  \\
		Te         & 2.71$\pm$0.07           & 0.53$\pm$0.08     &    \\
		\hline
                 \multicolumn{4}{l}{${\rm ^a}$ Average from EMIR and IGRINS results.}\\
                 \multicolumn{4}{l}{${\rm ^b}$ log(X/H)$_{PN}$ - log(X/H)$_{\odot}$. ${\rm ^c}$ Literature abundances}\\
                 \multicolumn{4}{l}{for NGC~7027 from \citet{sterlingetal16} and}\\
                 \multicolumn{4}{l}{\citet{sterlingetal17}; those for IC~418 are}\\
                 \multicolumn{4}{l}{from SPD15.}\\
\end{tabular}                 
\end{table} 		

\subsection{Comparison to Theoretical Models of AGB Evolution} \label{sec:mod}

Enrichments of  \emph{n}-capture elements have been predicted by several groups that model AGB evolution and nucleosynthesis. The resulting abundance patterns are similar for progenitor masses $M<4$--5~M$_{\odot}$, while the absolute enrichments tend to vary between different sets of models and even for different treatments of the physics by the same group. We have compared our results with the final abundances calculated by \citet{karakaslugaro16} and \citet{cristalloetal15} for solar metallicity (Z=0.014) AGB stars, since our recalculated O abundances are near the solar value (12 + log(O/H) = 8.69 for NGC\,7027, 8.62 for IC\,418). While the former models predict higher enrichments overall, in better agreement with our measurements, the relative enrichments of the observed elements are generally consistent with both sets of models for initial masses of 2 -- 4 M$_{\odot}$. Our observations agree with the predictions that the enrichments of Kr and Rb are larger than those of Se and Br. The enrichment of Te changes little ($\leq$~0.2 dex) over this mass range, while the elements Se--Rb show greater variations. We do not find Te to be enriched more strongly than Kr as predicted by the models, but this may be due to uncertainties in our approximate Te ICF. The enrichments for NGC\,7027 are consistent with a progenitor initial mass of around 3~$\pm$~0.5~M$_{\odot}$ as estimated by \citet{zhangetal05} and other studies \citep[see][for references]{sterlingetal16}. The Kr enrichment of IC\,418 suggests a lower mass for its progenitor star, which has been  estimated as M = 1.75 M$_{\odot}$\citep{morissetgeorgiev09} and as M = 1.4 $\pm$ 0.5 M$_{\odot}$\citep{henryetal18}.


\subsection{Abundances and Origins of Br and Te in Astronomical Sources} \label{sec:mod}

Br and Te are difficult to detect in stars, particularly in giants, since their strongest neutral transitions lie in the UV where red giants emit little flux. 

Large enhancements of Br have been reported in HgMn stars \citep[e.g.,][]{cowleywahlgren06} and He-rich DO white dwarfs \citep{werneretal18}, but these are attributed to chemical stratification due to diffusion and radiative levitation that result in photospheric abundances unrepresentative of overall composition.
Faint optical lines of Br$^{2+}$ have previously been reported in six PNe \citep[][SH07, MGR17]{pequignotbaluteau94}.  However, {\fbriii}~$\lambda$6130.37 can be blended with {\ciii}~$\lambda$6130.30 \citep{garciarojasetal15}, and the unrealistically large Br abundances derived from {\fbriii} $\lambda$6555.56 (SH07, MGR17) led MGR17 to suggest that this line may be contaminated by an unknown feature. While SH07 tentatively identified {\fbriv} $\lambda$7368.00 and $\lambda$9450.50 in NGC\,7027, they question these identifications owing to uncertainties in the continuum level and the possible presence of instrumental ghosts. Therefore, at present the NIR {\fbrv} line is the most reliable indicator of Br abundances in high-excitation nebulae.

Te has been detected in HgMn stars \citep{cowleyetal06}, DO white dwarfs \citep[e.g.,][]{rauchetal17, hoyeretal18}, and metal-poor halo stars \citep[e.g.,][]{roedereretal14}.  The Te abundances in the halo stars are consistent with a scaled solar \emph{r}-process distribution, indicating little or no \emph{s}-process enrichment.  
While approximately 80\% of the solar system Te abundance was formed in the \emph{r}-process \citep{bisterzoetal11}, $^{122}$Te, $^{123}$Te, and $^{124}$Te are $s$-only isotopes that are produced in low-mass AGB stars \citep{takahashietal16}.

Thus PNe represent the only viable means of studying Te in one of its sites of origin, with one intriguing exception.  \citet{smarttetal17} obtained optical and NIR data of the kilonova AT~2017gfo associated with the neutron star merger event GW170817, and tentatively identified lines of {\tei} in the early-time spectrum.  Unfortunately, the Te abundance could not be determined, due to the lack of reliable oscillator strengths and the fact that {\teii} is expected to be the dominant ion.  Nevertheless, this possible detection and that of lanthanides in neutron star mergers provide motivation to study the production of \emph{n}-capture elements in all their sites of origin, in order to better constrain models of galactic chemical evolution.



\acknowledgments

S.M. and J.G.R. acknowledge support from the Spanish Ministerio de Econom\'{\i}a y Competividad under projects AYA2015-65205-P and AYA2017-83383-P. M.A.B. and N.C.S. acknowledge support from NSF award AST-1412928, and H.L.D. from AST-1715332. J.G.R. acknowledges support from the Severo Ochoa excellence programme (SEV-2015-0548). This work used the Immersion Grating Infrared Spectrograph (IGRINS) developed by the University of Texas at Austin and the Korea Astronomy and Space Science Institute (KASI) with the financial support of NSF grant AST-1229522, the University of Texas at Austin, and the Korean GMT Project of KASI. 
This work has made use of the NASA Astrophysics Data System.




\newpage
\begin{thebibliography}{}

\bibitem[{{Asplund} {et~al.}(2009){Asplund}, {Grevesse}, {Sauval},
  \& {Scott}}]{asplundetal09}
{Asplund} M., {Grevesse}, N., {Sauval}, A.~J., \& {Scott}, P., 2009, ARA\&A, 47, 481

\bibitem[{{Badnell}(1986){Badnell}}]{badnell86}
{Badnell} N. R., 1986, J. Phys. B: At. Mol. Opt. Phys., 36, 4367

 \bibitem[{Badnell(1997)}]{bad97}
Badnell, N.~R.\ 1997, JPhB, 30, 1

\bibitem[{{Badnell}(2011){Badnell}}]{badnell11}
{Badnell} N. R., 1986, J. Computer Phys. Communications, 182, 7

\bibitem[{{Bautista} {et~al.}(1996){Bautista}, {Peng}, \& {Pradhan}}]{bautista96}
{Bautista} M.~A., {Peng} J., \& {Pradhan}, A., 1996, \apj, 460, 372

\bibitem[{{Bernard-Salas} {et~al.}(2001){Bernard Salas}, {Pottasch}, {Beintema, }\& {Wesselius}}]{bernardsalasetal01}
Bernard Salas, J., Pottasch, S.\ R., Beintema, D.\ A., \& Wesselius, P.\ R., 2001, \aap, 367, 949

\bibitem[{Berrington et al.(1987)}]{rmatrx}
Berrington, K.~A., Burke, P.~F., Butler, K., Seaton, M.~J., Storey, P.~J., Taylor, K.~T., \& Yu Yan\ 1997, J. Phys. B: At. Mol. Phys. 20, 6379

 \bibitem[\protect\citeauthoryear{Bisterzo et al.}{Bisterzo et~al}
{2011}]{bisterzoetal11}
Bisterzo, S., Gallino, R., Straneiro, O., Cristallo, S., \& K\"appeler, F., 2011, \mnras, 418, 284

 
\bibitem[\protect\citeauthoryear{{Busso}, {Gallino}, {Lambert}, {Travaglio} \&
  {Smith}}{{Busso} et~al.}{2001}]{bussoetal01}
{Busso} M.,  {Gallino} R.,  {Lambert} D.~L.,  {Travaglio} C.,    {Smith} V.~V.,
   2001, ApJ, 557, 802

 \bibitem[\protect\citeauthoryear{{Cowley} \& {Wahlgren}}{{Cowley}
  \& {Wahlgren}}{2006}]{cowleywahlgren06}
{Cowley} C.\ R.,  {Wahlgren} G.\ M., 2006, \aap, 447, 681

 \bibitem[\protect\citeauthoryear{Cowley et al.}{Cowley et~al}
{2006}]{cowleyetal06}
Cowley, C.\ R., Hubrig, S., Gonz\'alez, G.\ F., \& Nu\~nez, N., 2006, \aap, 455, L21

\bibitem[\protect\citeauthoryear{{Cristallo}, {Straniero}, {Piersanti} \&
{Gobrecht}}{{Cristallo} et~al.}{2015}]{cristalloetal15}
{Cristallo} S., {Straniero} O., {Piersanti} L., {Gobrecht} D., 2015, ApJS, 219, 40 

\bibitem[\protect\citeauthoryear{{Delgado-Inglada}, {Morisset} \&
  {Stasi{\'n}ska}}{{Delgado-Inglada} et~al.}{2014}]{delgadoingladaetal14}
{Delgado-Inglada} G.,  {Morisset} C.,    {Stasi{\'n}ska} G.,  2014, MNRAS, 440, 536, D-I14

 \bibitem[\protect\citeauthoryear{{Dinerstein}}{{Dinerstein}}{2001}]{dinerstein01}{Dinerstein} H.~L.,  2001, ApJL, 550, L223


\bibitem[\protect\citeauthoryear{{Garc{\'{\i}}a-Rojas},
  {Madonna}{Luridiana}{Sterling}{Morisset}{Delgado-Inglada} \& {Toribio San Cipriano}}{{Garc{\'{\i}}a-Rojas} et~al.}{2015}]{garciarojasetal15}
{Garc{\'{\i}}a-Rojas} J.,  {Madonna} S., {Luridiana} V., {Sterling} N. C., \,{Morisset} C., {Delgado-Inglada} G., {Toribio San Cipriano} L., 2015, MNRAS, 452, 2606

\bibitem[\protect\citeauthoryear{{Garz{\'{o}}n},
  {Abreu}{Barrera}{Becerril}{Cair{\'{\o}}s}{D{\'{\i}}az}{Fragoso}{Gago}{Grange}{Gonz{\'{a}}lez}{L{\'{o}}pez}{Patr{\'{o}}n}{P{\'{e}}rez}{Rasilla}{Redondo}{Restrepo}{Saavedra}{S{\'{a}}nchez}{Tenegi}\& {Vallb{\'{e}}}}{{Garz{\'{o}}n} et~al.}{2006}]{garzonetal06}
{Garz{\'{o}}n} F.,  {Abreu} D., {Barrera} S., {Becerril} S., {Cair{\'{o}}s} L.~M., {D{\'{\i}}az} J.~J., {Fragoso} A.~B., {Gago} F., {Grange} R., {Gonz{\'{a}}lez} C., {L{\'{o}}pez} P., {Patr{\'{o}}n} J., {P{\'{e}}rez} J., {Rasilla} J.~L., {Redondo} P., {Restrepo} R., {Saavedra} P., {S{\'{a}}nchez} V., {Tenegi} F., {Vallb{\'{e}}} M., 2006, Proc. of SPIE 6269, 18-7

\bibitem[\protect\citeauthoryear{{Garz{\'{o}}n},
  {Castro-Rodr{\'{\i}}guez}{Insausti}{L{\'{o}}pez-Mart{\'{\i}}n}{Hammersley}{Barreto}{Fern{\'{a}}ndez}{Joven}{L{\'{o}}pez}{Mato}{Moreno}{Nu\~{n}ez}{Patr{\'{o}}n}{Rasilla}{Redondo}{Rosich}{Pascual}\& {Grange}}{{Garz{\'{o}}n} et~al.}{2014}]{garzonetal14}
{Garz{\'{o}}n} F.,  {Castro-Rodr{\'{\i}}guez} N., {Insausti} M., {L{\'{o}}pez-Mart{\'{\i}}n} L., {Hammersley} P., {Barreto} M., {Fern{\'{a}}ndez} P., {Joven} E., {L{\'{o}}pez} P., {Mato} A., {Moreno} H., {Nu\~{n}ez} M., {Rasilla} J.~L., {Redondo} P., {Rosich} J., {Pascual} S., {Grange} R., 2014, Proc. of SPIE 9147

\bibitem[\protect\citeauthoryear{{Henry}, {Stephenson}, {Miller Bertolami}, {Kwitter} \& {Balick}}{{Henry}
  et~al.}{2018}]{henryetal18}
Henry, R. B. C., Stephenson, B. G., Miller Bertolami, M. M., Kwitter, K. B., Balick, B., 2018, MNRAS, 473, 241

\bibitem[\protect\citeauthoryear{{Hoyer}, {Rauch}, {Werner}, \& {Kruk}}{{Hoyer}
  et~al.}{2018}]{hoyeretal18}
Hoyer, D., Rauch, T., Werner, K., \& Kruk, J.\ W., 2018, \aap, 612, A62

\bibitem[\protect\citeauthoryear{{Joshi}, {Tauheed} \& {Davidson}}{{Joshi}
  et~al.}{1992}]{joshietal92}
{Joshi} Y. N., {Tauheed} A., {Davidson} I. G., 1992, Can. J. Phys., 70, 740

\bibitem[\protect\citeauthoryear{{Kaplan}, {Dinerstein}, {Oh}}{{Kaplan} et~al.}{2017}]{kaplanetal17} {Kaplan} K.~F., {Dinerstein} H.~L., {Oh} H., et al. 2017, ApJ, 838, 152

\bibitem[\protect\citeauthoryear{{K\"{a}ppeler}, {Gallino}, {Bisterzo} \& {Aoki}}{{K\"{a}ppeler}
  et~al.}{2011}]{kappeleretal11}
{K\"{a}ppeler} F.,  {Gallino} R.,  {Bisterzo} S.,  {Aoki} W.,  2011, RevModPhys, 83, 157 

\bibitem[\protect\citeauthoryear{{Karakas} \&
  {Lattanzio}}{{Karakas} \& {Lattanzio}}{2014}]{karakaslattanzio14}
{Karakas} A.~I., {Lattanzio} J.~C.,  2014, PASA, 31, 62

\bibitem[\protect\citeauthoryear{{Karakas} \&
  {Lugaro}}{{Karakas} \& {Lugaro}}{2016}]{karakaslugaro16}
  {Karakas} A.~I., {Lugaro} M.,  2016, ApJ, 825, 26
  
\bibitem[\protect\citeauthoryear{{Lugaro}, {Karakas}, {Pignatari} \& {Doherty}}{{Lugaro}
  et~al.}{2017}]{lugaroetal17}
{Lugaro} M.,  {Karakas} A.~I.,  {Pignatari} M.,  {Doherty} C.~L.,  2017, IAUS, 323, 86  

\bibitem[\protect\citeauthoryear{{Luridiana}, {Morisset} \& {Shaw}}{{Luridiana}
  et~al.}{2015}]{luridianaetal15}
{Luridiana} V.,  {Morisset} C.,    {Shaw} R.~A.,  2015, A\&A, 573, A42

\bibitem[\protect\citeauthoryear{{Madonna}, {Garc{\'{i}}a-Rojas}, {Sterling}, {Delgado-Inglada},
  {Mesa-Delgado}, {Luridiana}, {Roederer} \& {Mashburn}}{{Madonna}
  et~al.}{2017}]{madonnaetal17}
{Madonna} S.,  {Garc{\'{i}}a-Rojas} J.,  {Sterling} N.~C.,  {Delgado-Inglada} G.,  {Mesa-Delgado} A.,
  {Luridiana} V.,  {Roederer} I.~U., {Mashburn} A.~L.,  2017, MNRAS, 471, 1341 (MGR17)
  
  \bibitem[\protect\citeauthoryear{{Mashburn}, {Sterling}, {Madonna}, {Dinerstein}, {Roederer}, {Geballe}}{{Mashburn} et~al.}{2016}]{mashburnetal16} 
{Mashburn}, A.~L., {Sterling} N.~C., {Madonna} S., et al. 2016, ApJL, 831, L3

 \bibitem[\protect\citeauthoryear{{Moore}}{{Moore}}{1957}]{moore57}{Moore} C.\ E., 1957, 
  Atomic Energy Levels, Vol.\ III.  National Bureau of Standards Circular No.\ 467, Washington, DC

\bibitem[\protect\citeauthoryear{{Morisset} \&
  {Georgiev}}{{Morisset} \& {Georgiev}}{2009}]{morissetgeorgiev09}
  {Morisset} C., {Georgiev} L.,  2009, A\&A, 507, 1517
   
  
 \bibitem[\protect\citeauthoryear{{P{\'e}quignot} \& {Baluteau}}{{P{\'e}quignot}
  \& {Baluteau}}{1994}]{pequignotbaluteau94}
{P{\'e}quignot} D.,  {Baluteau} J.-P.,  1994, A\&A, 283, 593

\bibitem[\protect\citeauthoryear{{Park} {Jaffe} \& {Yuk}}{{Park}
  et~al.}{2014}]{parketal14}
{Park} C.,  {Jaffe} D.~T.,  {Yuk} I.-S., et al. 2014, proc. SPIE, 9147, 91471D

\bibitem[\protect\citeauthoryear{{Pottasch} {Bernard-Salas} {Beintema} \& {Faibelman}}{{Pottasch}
  et~al.}{2004}]{pottaschetal04}
{Pottasch} S. R.,  {Bernard-Salas} J.,  {Beintema} D. A.,  {Faibelman} W. A., 2004, A\&A, 423, 593

\bibitem[\protect\citeauthoryear{{Rauch} et al.}{{Rauch}
  et~al.}{2017}]{rauchetal17}
Rauch, T., Quinet, P., Kn\"orzer, M., Hoyer, D., Werner, K., Kruk, J.\ W., \& Demleitner, M., 2017,
  \aap, 606, A105

 \bibitem[\protect\citeauthoryear{{Riyaz} {Tauheed} \& {Rahimullah}}{{Riyaz}
  et~al.}{2014}]{riyazetal14}
{Riyaz} A.,  {Tauheed} A.,  {Rahimullah} K., 2014, JQSRT, 147, 86


\bibitem[\protect\citeauthoryear{{Roederer et~al.\ 2014}}{{Roederer}
  et~al.}{2014}]{roedereretal14}
Roederer, I.\ U., Schatz, H., Lawler, J.\ E., et al., 2014, \apj, 791, 32

\bibitem[\protect\citeauthoryear{{Sharpee et~al.\ 2003}}{{Sharpee}
  et~al.}{2003}]{sharpeeetal03}
{Sharpee} B.,  {Williams}, R., {Baldwin} J.~A., {van~Hoof} P.~A.~M., 2003, ApJS, 149, 157
  
\bibitem[\protect\citeauthoryear{{Sharpee}, {Zhang}, {Williams}, {Pellegrini},
  {Cavagnolo}, {Baldwin}, {Phillips} \& {Liu}}{{Sharpee}
  et~al.}{2007}]{sharpeeetal07}
{Sharpee} B.,  {Zhang} Y.,  {Williams} R.,  {Pellegrini} E.,  {Cavagnolo} K.,
  {Baldwin} J.~A.,  {Phillips} M.,    {Liu} X.-W.,  2007, ApJ, 659, 1265 (SH07)

\bibitem[\protect\citeauthoryear{Smartt et al.}{{Smartt}
  et~al.}{2017}]{smarttetal17}
Smartt, S.\ J., Chen, T.-W., Jerkstrand, A., et al., 2017, Nature, 551, 75

\bibitem[\protect\citeauthoryear{{Sterling} \& {Dinerstein}}{{Sterling} \&
  {Dinerstein}}{2008}]{sterlingdinerstein08}
{Sterling} N.~C.,  {Dinerstein} H.~L.,  2008, ApJS, 174, 158 

\bibitem[\protect\citeauthoryear{{Sterling}, {Dinerstein}, {Kaplan}, \& {Bautista}}{{Sterling}
  et~al.}{2016}]{sterlingetal16}
{Sterling} N.~C.,  {Dinerstein} H.~L.,  {Kaplan} K. F., {Bautista} M. A.,  2016, ApJ, 819, 9

\bibitem[\protect\citeauthoryear{{Sterling}, {Madonna}, {Butler}, {Garc\'\i a-Rojas}, {Mashburn}, {Luridiana} \& {Roederer}}{{Sterling}
  et~al.}{2017}]{sterlingetal17}
{Sterling} N.~C., {Madonna} S., {Butler} K. et. al,  2017, ApJL, 840, 80

\bibitem[\protect\citeauthoryear{{Sterling}, {Porter} \&
  {Dinerstein}}{{Sterling} et~al.}{2015}]{sterlingetal15}
{Sterling} N.~C.,  {Porter} R.~L.,  {Dinerstein} H.~L.,  2015, ApJS, 218, 25 (SPD15)

\bibitem[\protect\citeauthoryear{{Storey} \& {Hummer}}{{Storey} \&
  {Hummer}}{1995}]{storeyhummer95}
{Storey} P.~J.,  {Hummer} D.~G.,  1995, MNRAS, 272, 41

\bibitem[\protect\citeauthoryear{{Takahashi}, {Blaum} \&
  {Novikov}}{{Takahashi} et~al.}{2016}]{takahashietal16}
Takahashi, K., Blaum, K., \& Novikov, Yu., 2016, \apj, 819, 118

\bibitem[\protect\citeauthoryear{{Tauheed} \&
  {Naz}}{{Tauheed} \& {Naz}}{2011}]{tauheednaz11}
  {Tauheed} A., {Naz} A.,  2011, J. Korean Phys. Soc., 59, 2910
    
\bibitem[\protect\citeauthoryear{{Vassiliadis} \&
  {Wood}}{{Vassiliadis} \& {Wood}}{1994}]{vassiliadiswood94}
  {Vassiliadis} E., {Wood} P. R.,  1994, \apjs, 92, 125

\bibitem[\protect\citeauthoryear{{Werner}, {Rauch}, {Kn\"orzer}, \& {Kruk}}{{Werner}
  et~al.}{2018}]{werneretal18}
Werner, K., Rauch, T., Kn\"orzer, M., \& Kruk, J.\ W., 2018, \aap, in press (arxiv:1803.04809)


\bibitem[\protect\citeauthoryear{{Zhang}, {Liu}, {Luo}, {P{\'e}quignot} \& {Barlow}}{{Zhang} et~al.}{2005}]{zhangetal05}
{Zhang} Y., {Liu} X.-W., {Luo} S.-G., {P{\'e}quignot} D., {Barlow} M.~J.,  2005, A\&A, 442, 249


\end{thebibliography}
\end{document}